\documentclass[%
 preprint,
 amsmath,amssymb,
 aip,pop,
 floatfix,
]{revtex4-2}

\usepackage{xcolor}
\usepackage{graphicx}
\usepackage{dcolumn}
\usepackage{bm}
\usepackage{xcolor}

\usepackage[utf8]{inputenc}
\usepackage[T1]{fontenc}
\usepackage{mathptmx}
\usepackage{hyperref}
\begin{document}



\title{
Reconstructing magnetic deflections from sets of proton images using differential evolution
}

\author{Joseph M. Levesque}
\email{jmlevesque@lanl.gov}
\affiliation{P-2, Fundamental and Applied Physics, Los Alamos National Laboratory, Los Alamos, New Mexico 87545, USA}

\author{Lauren J. Beesley}
\affiliation{A-1, Information Systems and Modeling, Los Alamos National Laboratory, Los Alamos, New Mexico 87545, USA}

\date{\today}

\begin{abstract}

Proton imaging is a powerful technique for imaging electromagnetic fields within an experimental volume, in which spatial variations in proton fluence are a result of deflections to proton trajectories due to interaction with the fields.
When deflections are large, proton trajectories can overlap, and this nonlinearity creates regions of greatly increased proton fluence on the image, known as caustics.
The formation of caustics has been a persistent barrier to reconstructing the underlying fields from proton images.
We have developed a new method for reconstructing the path-integrated magnetic fields which begins to address the problem posed by caustics.
Our method uses multiple proton images of the same object, each image at a different energy, to fill in the information gaps and provide some uniqueness when reconstructing caustic features.
We use a differential evolution algorithm to iteratively estimate the underlying deflection function which accurately reproduces the observed proton fluence at multiple proton energies simultaneously.
We test this reconstruction method using synthetic proton images generated for three different, cylindrically symmetric field geometries at various field amplitudes and levels of proton statistics, and present reconstruction results from a set of experimental images.
The method we propose requires no assumption of deflection linearity and can reliably solve for fields underlying linear, nonlinear, and caustic proton image features for the selected geometries, and is shown to be fairly robust to noise in the input proton intensity.

\end{abstract}

\maketitle 

\section{Introduction}\label{sec:Introduction}

Reconstructing the underlying path-integrated magnetic fields from proton images is currently a topic of interest in magnetized high-energy-density (HED) experiments.
In these experiments, deflections of the probe protons due to interaction with electromagnetic fields produce observable features in images of the proton fluence.
The proton images only provide information of the final position of the protons at the image plane, and not their velocity.
Reconstructing the deflection map or path-integrated field from a proton image is therefore nontrivial.

Multiple methods exist for reconstructing path-integrated magnetic fields from proton images in which the deflections are small and the proton image intensity can be related to the deflections by a simple functional transformation.\cite{Levy_RSI2015,Graziani_RSI2017,Bott_JPP2017,Palmer_PoP2019}
Unfortunately, these existing methods tend to fail when the imparted proton deflections are such that neighboring proton trajectories intersect before reaching the image.
When proton trajectories intersect, nonlinear increases in proton fluence known as caustics are observed on the image.\cite{Kugland_RSI2012} 
Because of the nonlinearity in proton positions on the image with respect to initial trajectory, when caustics are present the simple functional transformation between proton image fluence and proton deflections breaks down.
Caustics are an important aspect of proton imaging, arising from sufficient combinations of field amplitude and field gradients, and are commonly observed in magnetized HED experiments.\cite{Kugland_NPhys2012,Kugland_PoP2013,Huntington_NP2015,Huntington_PoP2017}

Lacking an accurate transformation between the proton image and deflections, it has so far been challenging, if not impossible, to accurately reconstruct from a proton image the path-integrated magnetic field which produce caustic features, though some attempts have been made to bridge this gap. 
For example, \citet{Kasim_PRE2017} used a computational geometry approach to reconstruction, succeeding with some test images in the weakly caustic regime, but failing when more pronounced, branching caustics appeared on the image. 
\citet{Chen_PRE2017} trained a neural network to reconstruct a pseudo-3D magnetic field map from proton images for a limited set of field geometries, and also saw some success for some caustic test images, but failed in other instances as a result of insufficient information from the provided proton image.
Additionally, the performance of any neural network will depend heavily on its organization and the training set provided, which may reduce accuracy when presented with images of novel field configurations, which are likely to occur in experiment.

In this paper we introduce a new method of reconstructing underlying magnetic deflection fields from caustic proton images by using sets of proton images of the same field structure at different proton energies.
We use a differential evolution algorithm --- a type of genetic optimization algorithm --- to iteratively estimate the proton deflections as a function of initial proton trajectory and proton energy, searching for deflections which minimize a metaheuristic cost function, simultaneously minimizing the error between the synthetic reconstruction images and the input images in addition to an additional heuristic parameter which further constrains the solution space to guide the reconstruction toward more physically relevant solutions.
We restrict analysis and demonstration of this technique to magnetic fields in which the proton deflection is solely in the radial direction so that we can consider pseudo-one dimensional proton intensity profiles.
Our method of reconstructing deflections from a set of images, along with a realistic binning method to create the proton images, allows accurate reconstruction of deflections even in the caustic regime and further expands the regimes for which quantitative measurements of magnetic fields can be made via proton imaging.

Section \ref{sec:ProtonImaging} outlines some basics of proton imaging, illustrating the general imaging geometry and our assumptions of radial deflections.
Section \ref{sec:Method} describes our differential evolution reconstruction method.
Section \ref{sec:Results} displays results of this reconstruction method for three test cases: a Gaussian ellipsoid vector potential, a current-carrying wire segment, and a modified current-carrying wire segment with an additional Gaussian drop-off, each at three different field strengths.
We demonstrate the robustness of our reconstruction method for varying levels of proton statistics, as well as reconstruction from only a single proton image at a time, and also show the reconstruction from a pair of experimental images.
In Section \ref{sec:Discussion} we discuss the benefits of this reconstruction method, with some insight on how it may be extended to reconstruction of arbitrary two-dimensional proton images.
Section \ref{sec:Conclusion} concludes the paper.

\section{Proton Imaging}\label{sec:ProtonImaging}

\begin{figure}[ht]
	\centering
	\includegraphics{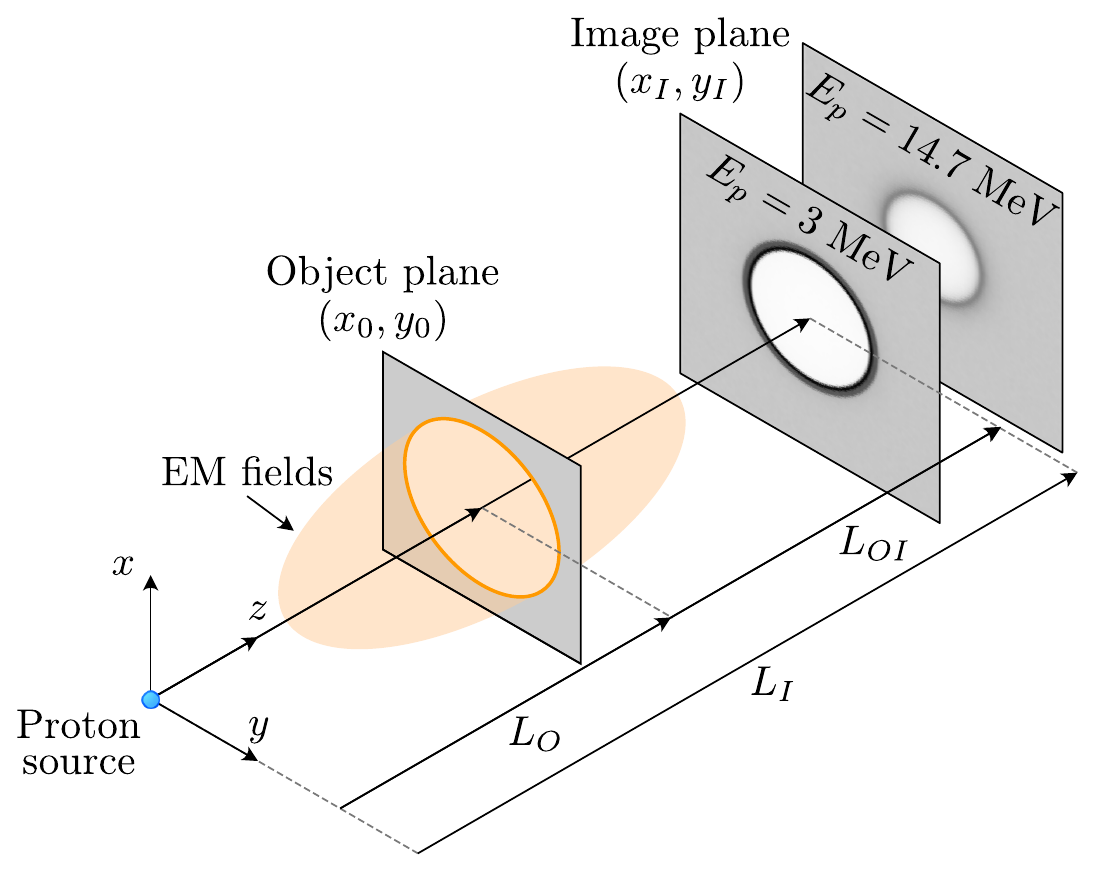}
	\caption{
	Illustration of the general proton imaging geometry (not to scale).
	The protons are initialized from a point source and probe primarily along the $z$ axis.
	The object plane is defined as the center of the interaction region, and the image plane is where the images are generated.
	}
	\label{fig:ProtonImaging}
\end{figure}

In this section we describe the process of proton imaging.
The general proton imaging geometry is illustrated in Figure \ref{fig:ProtonImaging}, in which the axis from the proton source to the object and image planes is $z$, $L_O$ is the distance from the source to the center of the interaction region (the object plane), $L_I$ is the distance from the proton source to the image plane, and $L_{OI}$ is the distance from the object plane to the image plane.
We limit our consideration to fields that deflect protons only in the outward radial direction.
Proton deflections by magnetic fields in the interaction region are a function of the perpendicular magnetic field integrated along the proton's trajectory and the energy of the proton, as
\begin{equation}\label{eq:alpha}
	\Delta\alpha(x_0,y_0) = \frac{q}{\sqrt{2 m E_p}} \int B_{\phi}(x_0,y_0) \text{d}l,
\end{equation}
where $\Delta\alpha$ is the deflection angle, $q$ is the proton charge, $m$ is the proton rest mass, $E_p$ is the kinetic energy of the proton, $B_{\phi}$ is the magnetic field of interest, and $l$ represents the path of the proton through the interaction region.
We assume that the protons are emitted from a point source, so the protons naturally diverge with initial trajectories defined by the angles $\alpha_0$.

To simplify the imaging process, we also assume that the deflections are only important over distances greater than the size of the interaction region, and that the trajectory of any proton remains constant within the interaction region.
Using these assumptions, we calculate the deflection from the total path-integrated field along any ballistic trajectory and apply the deflection in a single step, similarly to the approach of \citet{Levesque_PoP2019}, backtracking to add the deflection at the object plane.
Recognizing that the deflections are limited to the radial direction, the deflection and final position of a proton of energy $E_p$ only depend on the initial trajectory $\alpha_0$, and we can rewrite equation \ref{eq:alpha} as
\begin{equation}\label{eq:Dalpha}
	\Delta\alpha(\alpha_0) = \frac{q}{\sqrt{2 m E_p}} \int B_{\phi}(\alpha_0) \text{d}l.
\end{equation}
In this form, the location at the image plane of any proton deflected by an amount $\Delta\alpha$ to its final trajectory $\alpha_0+\Delta\alpha$ can be calculated as
\begin{align}\label{eq:rI}
	r_I(\alpha_0) &= L_{OI}\tan(\alpha_0+\Delta\alpha(\alpha_0)) +L_O\tan(\alpha_0)
\end{align}
where the first term is the radial distance travelled along the trajectory from the object plane to the image plane, and the second term is the distance travelled along the initial trajectory from the source to the object plane.
When the fields are strong enough to significantly alter the proton trajectories within the interaction region, deflecting protons based on the total path-integrated fields begins to break down, and a particle tracking method should be used instead, to project protons through the interaction region in a way which incorporates time-varying deflections, though incurring additional computational cost.

When deflections are small and protons along neighboring trajectories do not cross before reaching the image plane, it is usually sufficient to represent the image resulting from the deflected protons by a simple transformation depending on the Jacobian of the deflection in the image plane as
\begin{equation}\label{eq:ImageTransform}
I = \frac{I_0+\delta}{|(r_I/r_0)\partial r_I/\partial r_0}|+\delta
\end{equation}
where $I$ is the proton fluence at the image, $I_0$ is the unperturbed proton fluence at the object plane, the coordinates $r_I$ and $r_0$ correspond to positions at the image plane and the object plane, respectively, and $\delta$ is a small correction factor to limit the intensity.\cite{Kugland_RSI2012}
Existing field reconstruction methods are mostly limited to this small-deflection regime and reconstruct the field from single proton images by solving the Poisson equation or an optimal transport criterion.\cite{Graziani_RSI2017,Bott_JPP2017}
When protons are deflected in such a way that trajectories intersect one another, caustics appear on the image and the image transformation of equation (\ref{eq:ImageTransform}) breaks down because the final location of protons on an image as a function of initial trajectory is no longer single-valued.
In terms of equation (\ref{eq:ImageTransform}), caustics appear where the denominator goes to zero, which predicts very high (effectively infinite) proton intensity, limited only by the chosen value of $\delta$, though changing $\delta$ to limit the intensity jump also changes the location of the caustic.
When this transformation breaks down, existing methods of reconstruction typically also break down, though some reconstruction methods have proposed a possible extension of the transformation into the caustic regime.\cite{Kasim_PRE2017,Bott_JPP2017}

In this paper we are concerned almost exclusively with the caustic regime, which requires that we fully consider the nonlinear proton trajectories.
When generating synthetic proton images we account for the deflection of each proton, propagate them through the system, and bin them into discrete pixels at the image plane, mimicking the physical proton imaging procedure.
Aside from the more complicated imaging process, the real difficulty in any reconstruction is that we only have a measure of the captured proton fluence and not proton velocity. 
The nonlinearity of proton trajectories introduced by caustics requires a nonlinear reconstruction approach.

\section{Reconstruction Method}\label{sec:Method}

\begin{figure}[ht]
	\centering
	\includegraphics{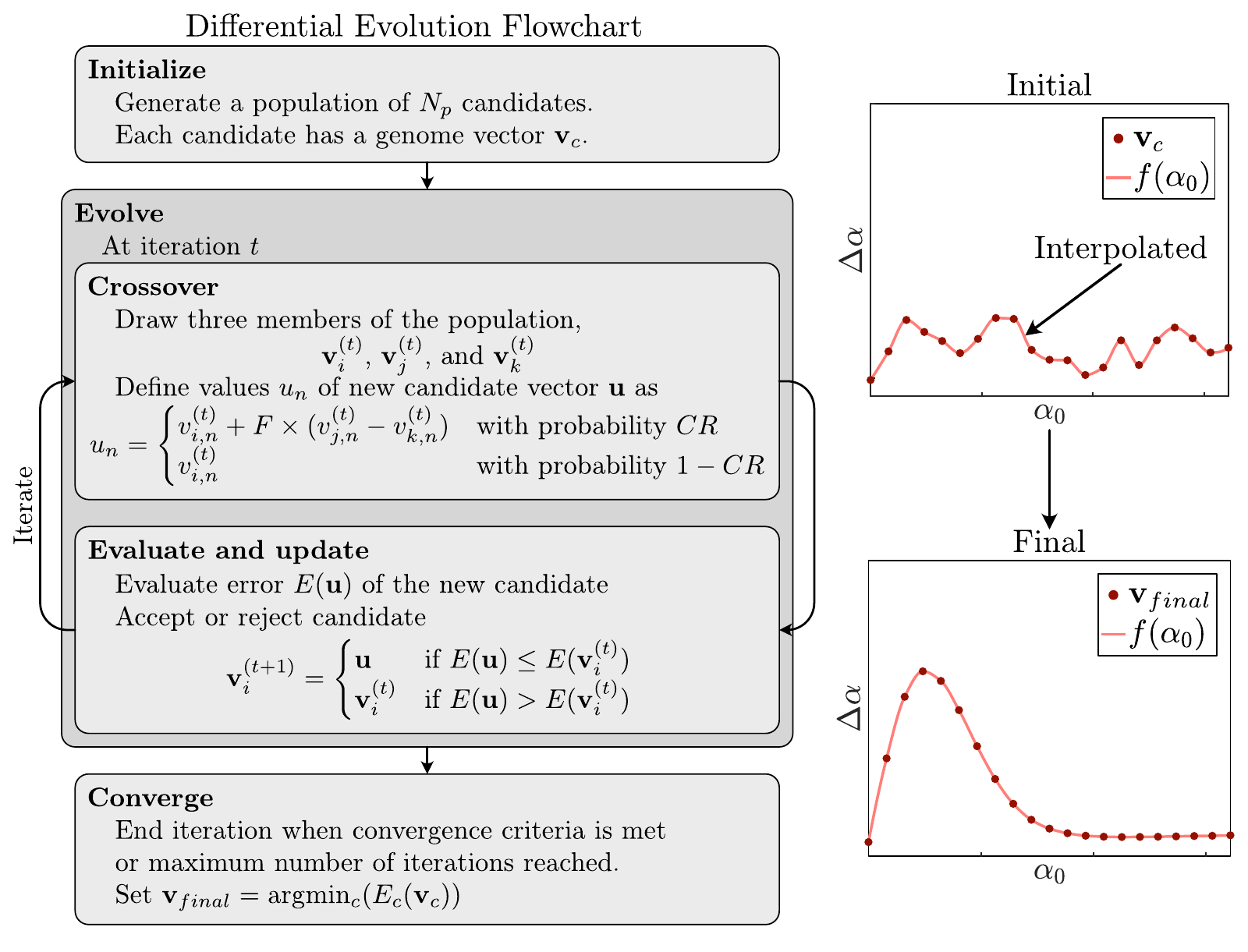}
	\caption{
	A flowchart illustrating the general differential evolution algorithm, with examples of an initial candidate and the final reconstruction, showing the deflection values of $\mathbf{v}_c$ at discrete nodes and the continuous deflections $f(\alpha_0)$.
	}
	\label{fig:Flowchart}
\end{figure}

So how exactly can we hope to reconstruct deflection profiles which create caustic features if there is no single-valued mapping of intensity?
We make use of an overlooked feature of most proton imaging experiments --- multiple images of the same object at different proton energies.
Recall from equation (\ref{eq:alpha}) that for a given path-integrated magnetic field, protons are deflected proportionally to their energy as $E_p^{-1/2}$.
Importantly, however, the location of image features does \emph{not} share this proportionality.
Therefore, by recognizing the lack of correlation between image features and deflections we can use multiple images of the same field to provide additional information and achieve some uniqueness when reconstructing the deflections.

We base our imaging on experiments at the OMEGA laser facility, a user facility with 60 independent, high-power lasers.\cite{Boehly_RSI1995} 
Proton imaging experiments at this facility commonly use the implosion of a D$^3$He capsule to produce a point-source of quasi-monoenergetic protons at 3 MeV and 14.7 MeV.\cite{Li_RSI2006,Li_PoP2009}
We now redefine equation (\ref{eq:alpha}) to the more general form
\begin{equation}\label{eq:alphaF}
	\Delta\alpha(\alpha_0,E_p) = f(\alpha_0)/\sqrt{E_p},
\end{equation}
where the deflection angle for any proton trajectory $\alpha_0$ is only a function of some value $f$ and the proton energy.
The goal of our reconstruction method is to find the underlying deflection field $f$ which minimizes the error between synthetic, reconstructed images and the input images for the chosen proton energies simultaneously, in addition to minimizing extra heuristics.

In reconstructing the underlying deflection field of a caustic feature, we essentially need to determine the trajectory of each proton from its origin to the image to account for intersecting trajectories.
Because of the large number of incident protons $N_p$ in any image, solving for each proton trajectory individually --- a continuous $N_p$-dimensional space --- is computationally prohibitive.
To more easily traverse the space of possible solutions we instead define an estimated $\Delta\alpha_n$ at a number $n$ of nodes in space, where the nodes are fixed in $\alpha_0$ position, from which we create a continuous deflection field $f(\alpha_0)$ by interpolating $\Delta\alpha_n$ between the node points as illustrated on the right side of Figure \ref{fig:Flowchart}.
Using the continuous $\Delta\alpha$ we create the continuous deflected trajectories $\alpha_0+\Delta\alpha(\alpha_0,E_p)$ and generate synthetic images for arbitrary initial proton distributions in $\alpha_0$ for the desired proton energies $E_p$.
This method reduces the dimensionality of the parameter space to $n$ dimensions, with $n$ only on the order of 10, and greatly reduces computational cost.
Even with a greatly reduced number of free parameters, there is still a large solution space to explore, so we need a method which allows us to quickly traverse the space to find a deflection function which best reproduces the images for our desired constraints.


We chose to solve this optimization problem using differential evolution, a type of of genetic algorithm.\cite{Storn_1997,Das_SwEvo2016}
Differential evolution (DE) is an iterative optimization method designed to handle difficult, high-dimensional, non-convex problems.
DE works by iteratively updating a population of solution candidates to find more optimal solutions based on the applied heuristic cost function.
Figure \ref{fig:Flowchart} illustrates the general DE procedure, which can generally be described in three steps: initialization, evolution, and convergence.
Following initialization, each candidate $c$ within the population has its own genome, in this case a vector $\mathbf{v}_c$ defining the deflections at each node $n$, where ${v}_{c,n} = \Delta\alpha_{c,n}$, from which synthetic proton images at both energies are generated, and the candidate is scored based on its fitness.
 
For the evolution process, we base our DE implementation on the canonical ``DE/Rand/1'' algorithm\cite{Das_SwEvo2016}, where at every iteration $t$ the algorithm creates a new, offspring candidate vector $\mathbf{u}$ by altering a number of randomly selected nodes N via the crossover process.
During evolution, the crossover rate $CR$ --- a user-defined value between 0 and 1 --- defines the probability that any given node will undergo crossover, or a modification of it's genome, at each iteration.
The nodes that undergo crossover are determined by drawing $n$ random values between 0 and 1, where the nodes $N$ which drew a value less than $CR$ participate in the crossover step.
The crossover values of a target candidate $i$ are calculated from the node values of two other randomly selected parent candidates $j$ and $k$, as
\begin{align}
	u_n = 
	\begin{cases}
    	v_{i,n}^{(t)}+F \times (v_{j,n}^{(t)}-v_{k,n}^{(t)})		& \text{if } n \in N\\
    	v_{i,n}^{(t)}           & \text{otherwise}
	\end{cases}
\end{align}
where $F$ --- another user-defined value between 0 and 2 --- is the scaling parameter, and determines the amount of change.
$CR$ and $F$ are the primary tuning parameters for DE, and affect the stiffness and rate of convergence of the population. 
For example, as the crossover rate increases the average number of nodes during crossover increases.
Unlike other genetic algorithms, the population size in DE remains static throughout --- candidate vectors are simply updated with better genomes as they are found.
If the deflection function of the offspring candidate produces images with a lower error than the original target candidate, the target candidate takes on the values of the offspring vector, $\mathbf{v}_i^{(t+1)} = \mathbf{u}$, and if the error is greater, the offspring candidate is rejected and $\mathbf{v}_i^{(t+1)} = \mathbf{v}_i^{(t)}$. 
The algorithm repeats until some stopping criteria for the error are met.

By using a population of candidates that can be distributed over a larger solution space, and because it can alter many values per iteration, DE is able to quickly explore a solution space and select more optimal solutions without getting stuck in shallow local minima.
When dealing with high-dimensional, non-convex systems, simpler optimization methods like a strict gradient descent algorithm will undoubtedly get stuck in various local minima.
More complicated optimization algorithms like stochastic gradient descent with batching can likely produce similar results to our DE method, but are complicated and would likely require different means of tuning.
DE is relatively simple to implement, can be initialized with varying levels of complexity, and converges quickly.

In our specific implementation of DE, we set the number of candidates $N_c$ in our population to only 100 to improve convergence since we have $n \lesssim 30$, as suggested by \citet{Piotrowski_SwEvo2017}.
Higher-dimensional systems would instead require between $3n$ and $5n$ candidates to have a sufficiently large population for convergence.
We are only considering one-dimensional radial deflections, so in our DE algorithm for reconstructing deflections the genome vector $\mathbf{v_c}$ of each candidate defines proton deflections $\Delta\alpha$ at the discrete node points.
We limit the total range of $\Delta\alpha$ to between 0 and 0.5.
Further limiting $\Delta\alpha$ to a smaller range --- between 0 and 0.05 for our purposes --- during initialization further improves the rate of convergence, by allowing the deflection field to essentially build itself up in the early iterations, and usually prevents large variations between neighboring nodes that can compound and result in solutions with large oscillations.
To create a continuous deflection function $f(\alpha_0)$ for each candidate we interpolate the genome $\mathbf{v_c}$ across the range of $\alpha_0$ that reach the image using the piecewise Hermite interpolating polynomial method (PCHIP), which allows for curvature in the interpolant while maintaining the overall shape.
We use $f(\alpha_0)$ to generate the synthetic pseudo-1D proton images $I_{1,\text{recon}}$ and $I_{2,\text{recon}}$ for each candidate at the desired proton energies.
The synthetic images are normalized to maintain the same relative number of protons as in the data images.
We next calculate the fitness of each candidate based on its synthetic images and deflection field using a number of metaheuristics. 

When calculating the fitness of a candidate, the primary error heuristic we want to minimize is based on the difference of the reconstructed images to the input images. 
However, to further reduce the space of possible solutions and increase the accuracy of the solutions we also add an additional heuristic which enforces some desired physical constraints.
The total error of each candidate is calculated using three error heuristics and their associated weights as
\begin{equation}\label{eq:Error}
	E_{c}(\mathbf{v}_c)^{(t)} = (w_1 \epsilon_1(\mathbf{v}_c)^{(t)}+w_2 \epsilon_2(\mathbf{v}_c)^{(t)})(1 + w_3 \epsilon_3(\mathbf{v}_c)^{(t)}),
\end{equation}
in which $E_c^{(t)}$ is the total heuristic error of candidate $c$ at evolution iteration $t$, and $w_i$ are scalar weights of the heuristics $\epsilon_i$ that act as tuning parameters for each heuristic.
The heuristics are calculated as follows:
\begin{align}
	\epsilon_1(\mathbf{v}_c) &= \text{mean}_r(|I_{1,\text{recon}}(\mathbf{v}_c)^2-I_1^2|/(I_1+\delta)^2), \label{eq:E1}\\
	\epsilon_2(\mathbf{v}_c) &= \text{mean}_r(|I_{2,\text{recon}}(\mathbf{v}_c)^2-I_2^2|/(I_2+\delta)^2), \label{eq:E2}\\
	\epsilon_3(\mathbf{v}_c) &= \int_{\alpha_{0,min}}^{\alpha_{0,max}} \left[\frac{\sqrt{1+\left(\partial^2 \Delta\alpha_c(\alpha_0,E_1)/ \partial\alpha_0^2 \right)^2}}{\alpha_{0,max}-\alpha_{0,min}}\right] \partial\alpha_0 -1. \label{eq:E3}
\end{align}
$\epsilon_1$ and $\epsilon_2$ are the primary heuristics, and are the mean of absolute square error between the candidate $c$'s reconstructed images $I_{1,\text{recon}}$ and $I_{2,\text{recon}}$, and the input images $I_1$ and $I_2$, normalized by the square of the input image intensity plus a small factor $\delta$, where the mean is taken over all points $r$ on the image, in which the subscripts 1 and 2 refer to the two proton energies. 
In the denominator of $\epsilon_1$ and $\epsilon_2$, $\delta=0.02$ is a small correction factor to prevent the error from approaching infinity where the image intensity is zero.
The summation of the image error at both energies ensures that the error of both (or all) images will be simultaneously minimized.
Because the image error is the primary value we want to minimize, the first image heuristic term on the left side of equation (\ref{eq:Error}) is allowed to go to zero, ensuring that this will be the most important heuristic throughout the search.
$\epsilon_3$ is the difference of the arc length of the gradient of the continuous proton deflections $\Delta\alpha_c(\alpha_0,E_1)$ of candidate $c$ at the lower proton energy $E_1=3$ MeV from the minimum arc length $(\alpha_{0,max}-\alpha_{0,min})$, normalized by the minimum length.
The heuristic $(1+w_3\epsilon_3)$ enforces a level of smoothness in both the gradient of the deflection field and the deflection field itself by effectively penalizing unnecessary oscillations between consecutive nodes in the gradient of the deflection field.
Limiting this secondary heuristic to a minimum value of 1 ensures that it will not dominate the reconstruction, particularly at later iterations.

For our tests, we typically use the following weights: $w_1=1$, $w_2=2$, and $w_3=5$.
The increased weight on the high-energy proton image error generally improves convergence, as the high-energy images are typically harder to fit.
These weights can be tuned to be more or less constraining ---  increasing $w_1$ will put more emphasis on fitting the lower-energy image, and reducing $w_3$ will reduce the smoothness constraint, for example.
Taken together, the total heuristic of equation \ref{eq:Error} with these weights define the fitness landscape of the solution space, and determine how the candidates converge.

The crossover and evaluation steps in our implementation are modified slightly from the standard ``DE/Rand/1'' algorithm in how the offspring vectors are generated, and in how they are accepted.
During crossover, the scaling factor $F$ is changed to $F(1+\text{rand})$ to allow for additional variation in the potential search space.
Additionally, even if the error of the offspring candidate is greater than the target candidate, it can still be accepted in our method.
We also occasionally allow the error to increase --- if the error is within 25\% of the original, there is a 20\% chance that the offspring will be accepted, although a higher-error offspring will always be rejected if the target candidate is currently the lowest-error candidate in the population.
These modifications are not strictly necessary, but the randomized scaling factor and the added chance of acceptance for the sub-optimal candidates in the population broadens the space available for mutations, and provides more ways for the population to escape from local minima.

The differential evolution algorithm proceeds using $CR=0.6$ and $F=0.5$, until the error reaches a certain threshold, at which point the candidates are reinitialized with $2n-1$ nodes, the starting values of which are obtained by pchip interpolation of the existing candidates.
The higher-resolution reinitialization allows finer alteration of the deflection field to further minimize error, continuing with $CR=0.2$.
Despite increasing the resolution, initializing a larger number candidates is not necessary for convergence because the lower-resolution population has already converged significantly.

\section{Results}\label{sec:Results}

In this section we test our reconstruction method on three cylindrically symmetric field configurations: a gaussian ellipsoid, a current-carrying wire, and a current-carrying wire with an additional exponential decrease.
Each of the magnetic fields in our test cases are defined in a three-dimensional functional form.
We calculate the total deflection field for each case by first integrating the magnetic field on a fine, regular grid within the range of initial trajectories that reach the image from the point source.
The path-integrated magnetic field map is converted into a deflection map for both 3 MeV and 14.7 MeV proton energies, corresponding to energies from the D-$^3$He capsule implosion source. 
To speed up calculation, we interpolate the discrete deflection maps create higher-resolution maps as a function of initial proton trajectory.
Interpolating the deflection map allows us to calculate deflections for arbitrary proton trajectories without having to perform integration for every proton trajectory.

To calculate the resulting images of these fields the protons are initialized as a point source in a uniform random distribution of $\alpha_0$ that reach the image plane.
The deflections are added at the center of the object plane $L_O$, and the protons travel with the new trajectories to the image plane at $L_I$ according to equation (\ref{eq:rI}). 
The two images are then created by binning the protons into discrete pixels in a $400\times400$ pixel image.
We typically initialize the images with $2\times 10^7$ protons to produce smooth intensity profiles with low Poisson noise, to which we add a $\sigma=2$ pixel Gaussian blur to the images to more closely resemble the fidelity of experimental CR39\cite{Seguin_RSI2003,Seguin_RSI2004} images.
Because we are dealing with cylindrically symmetric deflections, we reduce each two-dimensional image into an averaged radial proton intensity 200 pixels in length.
We use the one-dimensional proton fluence profiles as input to the reconstruction algorithm.

\begin{figure}[t]
	\centering
	\includegraphics{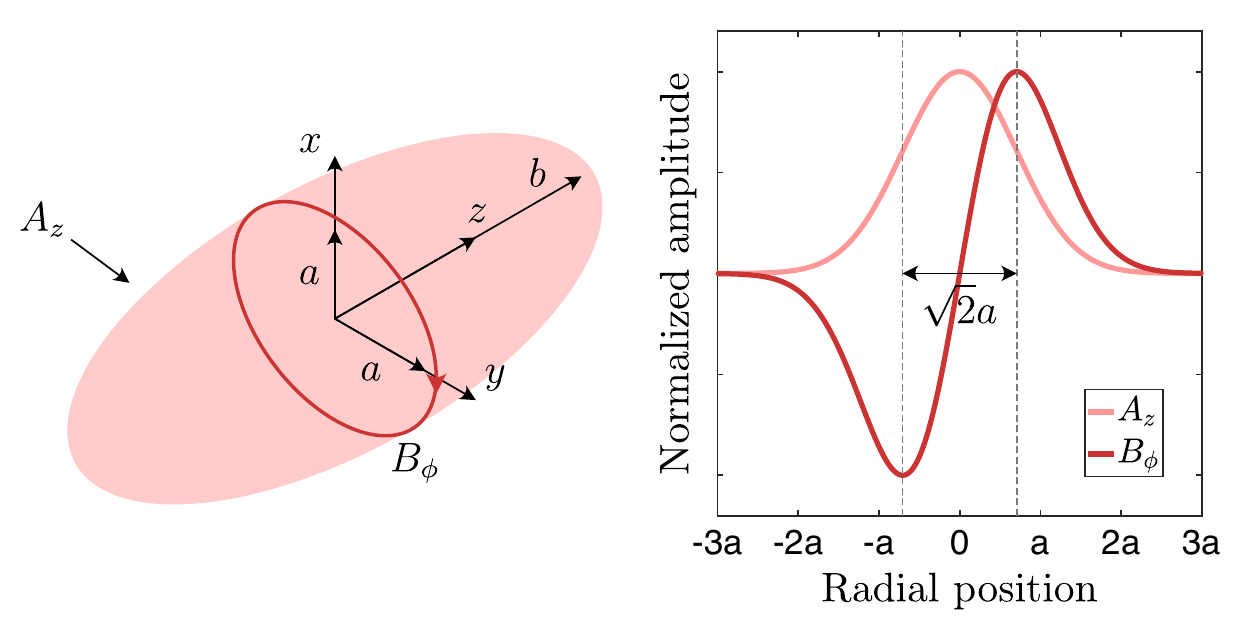}
	\caption{
	An illustration of the gaussian ellipsoid vector potential and corresponding magnetic field alongside normalized plots of these quantities as a function of radius at the center plane.
	The primary proton probe trajectory is along $z$, the same axis as the vector potential component.
	}
	\label{fig:EllipsoidField}
\end{figure}

\subsection{A Gaussian ellipsoid vector potential}

\begin{figure}[t]
	\centering
	\includegraphics{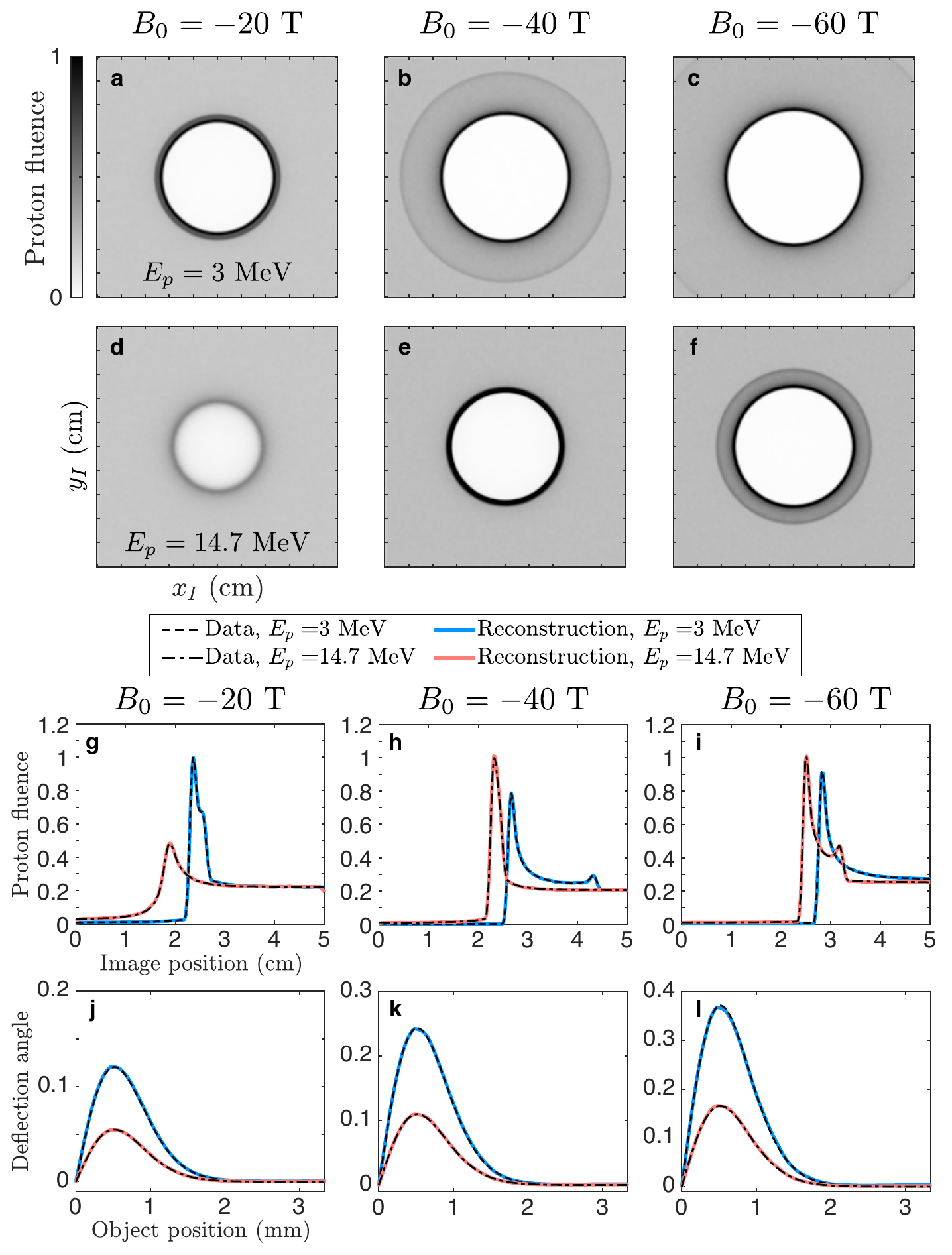}
	\caption{
	Synthetic test proton images and reconstruction results from the simulated Gaussian ellipsoid vector potential field as described by equation \ref{eq:GaussianEllipsoid}, using $a=750$ $\mu$m, $b=2$ mm, at three magnetic field amplitudes. 
	(a,b,c) Synthetic 3 MeV proton images.
	(d,e,f) Synthetic 14.7 MeV proton images.
	The proton images are scaled by the maximum fluence between both energy levels.
	(g,h,i) The known proton intensity (dashed lines) and reconstructed proton intensity profiles (solid lines).
	(j,k,l) The known radial deflection field (dashed lines) and reconstructed deflections (solid lines).
	}
	\label{fig:EllipsoidRecon}
\end{figure}

\begin{figure}
	\centering
	\includegraphics{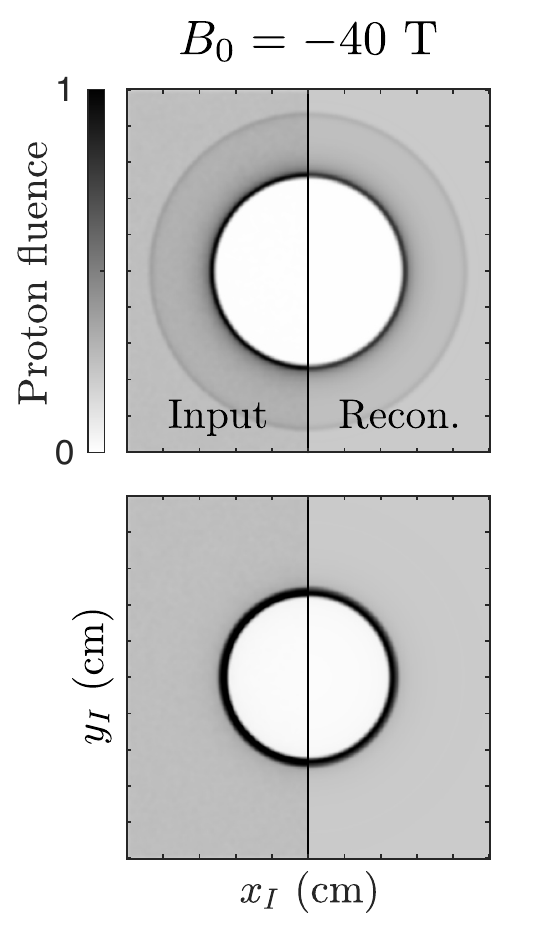}
	\caption{
	Comparison of (left) the input images and (right) reconstructed images extrapolated from the 1D reconstruction of Figure \ref{fig:EllipsoidRecon} for the -40 T field ellipsoid at (top) 3 MeV and (bottom) 14.7 MeV proton energies.
	}
	\label{fig:EllipsoidReconImage}
\end{figure}

Our first test case is the magnetic field defined by a single-component, Gaussian ellipsoidal vector potential.
The vector potential is defined as in \cite{Kugland_RSI2012},
\begin{equation}\label{eq:GaussianEllipsoid}
	A_z(x,y,z) = A_0 \exp\left[ -\frac{(x-x_c)^2}{a^2} -\frac{(y-y_c)^2}{a^2} -\frac{(z-z_c)^2}{b^2}\right],
\end{equation}
where $A_0$ is a constant that determines the amplitude of the vector potential, $(x_c,y_c,z_c)$ denotes the center of the ellipsoid, $a$ determines the effective radius of the ellipsoid perpendicular to the direction of the vector potential, and $b$ is the effective size of the ellipsoid along the direction of the vector potential.
For an ellipsoid centered at $x_c=y_c=0$, $r=\sqrt{x^2+y^2}$, the magnetic field corresponding to this vector potential is 
\begin{equation}
	B_{\phi}(r,z) = B_0\frac{r}{a}\exp\left[ -\frac{r^2}{a^2}-\frac{(z-z_c)^2}{b^2} \right],
\end{equation}
where $B_0$ is the magnetic field scaling factor which we use to define the field strengths.
The relationship between the vector potential and the magnetic field is illustrated in Figure \ref{fig:EllipsoidField}.

We only consider probing parallel to the axis of the vector potential, so the axis from the proton source to the image plane is $z$. 
Given the radial symmetry of the vector potential system and the orientation of $A_z$ and $B_{\phi}$ with respect to the proton trajectories, the protons will be deflected radially inward when $A_z$ and $B_{\phi}$ are positive, and radially outward when they are negative.
For simplicity we further restrict the analysis to outward proton deflections, so the vector potential and magnetic field in the following examples are negative.

Figure \ref{fig:EllipsoidRecon}(a-f) shows synthetic test images at 3 MeV and 14.7 MeV proton energies for the Gaussian ellipsoid system defined by $a=750$ $\mu$m, $b=2$ mm, at the magnetic field amplitudes $B_0=-20$ T, $-40$ T, and $-60$ T.
This range of field amplitudes demonstrates the increasing prevalence of caustics as deflections increase.
At $-20$ T, the 3 MeV image is in a weakly caustic regime, forming a void of protons in the center bounded by a branched, high-fluence ring; the 14.7 MeV image shows much less pronounced features and is in the linear deflection regime.
As the field increases to $-40$ T the features are much more distinct ---  the secondary peak of the low-energy image has greatly expanding in radius, and the high-energy image is entering the caustic regime.
At $-60$ T the protons which form the second peak in the low-energy image fall mostly outside of the image; the high-energy image is clearly caustic.
We use averaged radial lineouts from both images at each field amplitude as input to our reconstruction algorithm.

Figure \ref{fig:EllipsoidRecon}(g-i) shows the synthetic proton fluence profiles and the corresponding reconstructed proton intensity profiles, and Figure \ref{fig:EllipsoidRecon}(j-l) shows the actual deflection field alongside the reconstruction.
The reconstruction genomes are initialized with $n=21$ nodes to begin evolution, which increases to $n=41$ after reaching $\epsilon_1+\epsilon_2<0.25$, after which the computation proceeds for at most 300,000 iterations or until $\epsilon_1+\epsilon_2<0.01$.
Our reconstruction method very accurately reconstructs both the proton fluence profiles for both energies and the underlying deflection function.
The error in the final reconstructed proton images is almost negligible, less than a total deviation of 4\% from the input images across both images combined.
Figure \ref{fig:EllipsoidReconImage} compares the 2D test image with a 2D image extrapolated from the reconstructed 1D proton fluence profile at $-40$ T, further demonstrating that there is little difference between the two.
For the Gaussian ellipsoid system the reconstruction method works equally well regardless of whether the images are in the linear or caustic regimes, and is capable of reconstructing the deflections even when some information has been lost from the low-energy image at $B_0=-60$ T.

\begin{figure}
	\centering
	\includegraphics{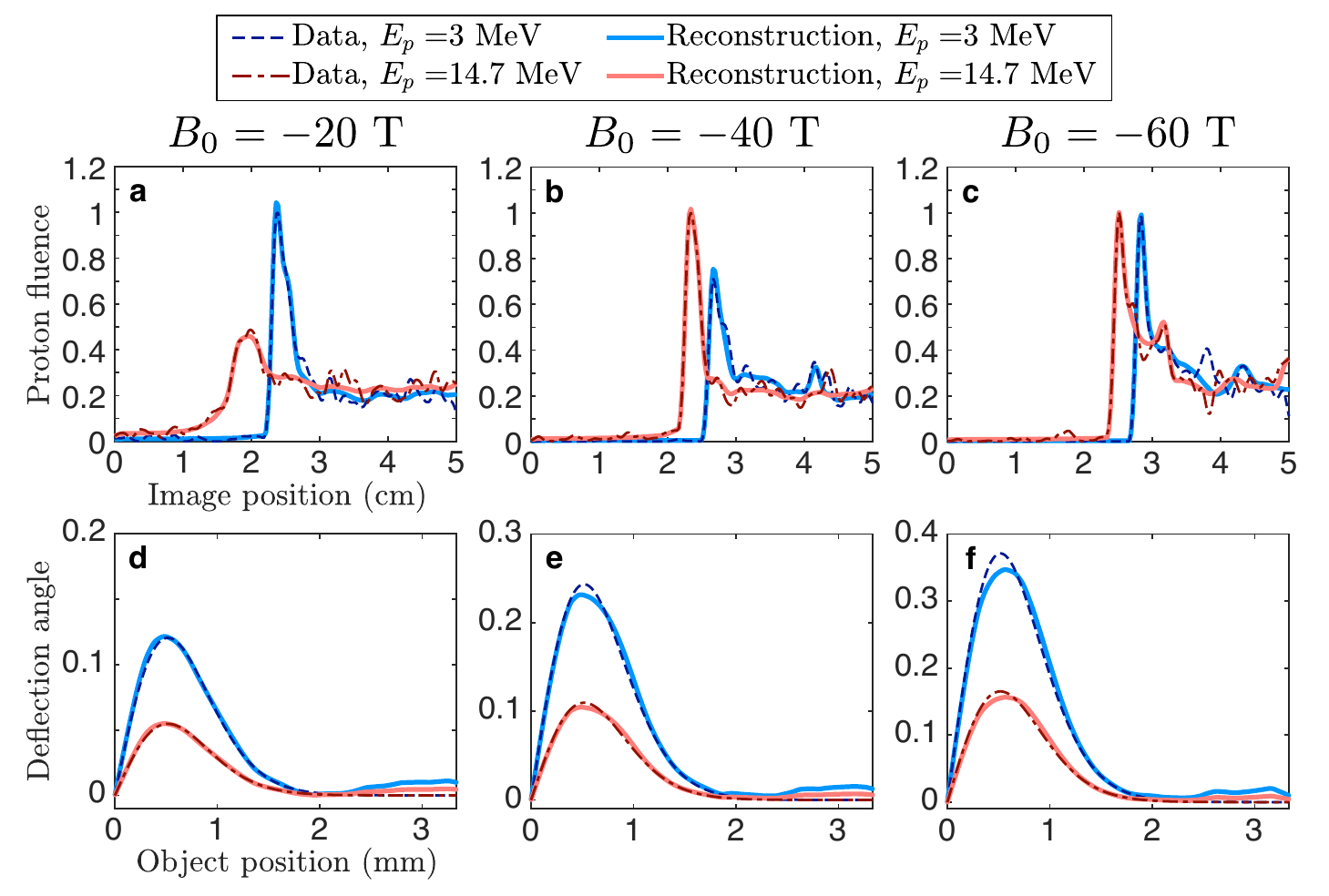}
	\caption{
	Results of our reconstruction method for proton images produced by the Gaussian ellipsoidal magnetic field with fewer protons per image, causing Poisson noise.
	(a,b,c) The known proton intensity (dashed lines) and reconstructed proton intensity profiles (solid lines).
	(d,e,f) The known radial deflection field (dashed lines) and reconstructed deflections (solid lines).
	The reconstruction method is able to reconstruct the deflection well despite the noisy intensity profiles.
	}
	\label{fig:EllipsoidReconNoise}
\end{figure}

\begin{figure}
	\centering
	\includegraphics{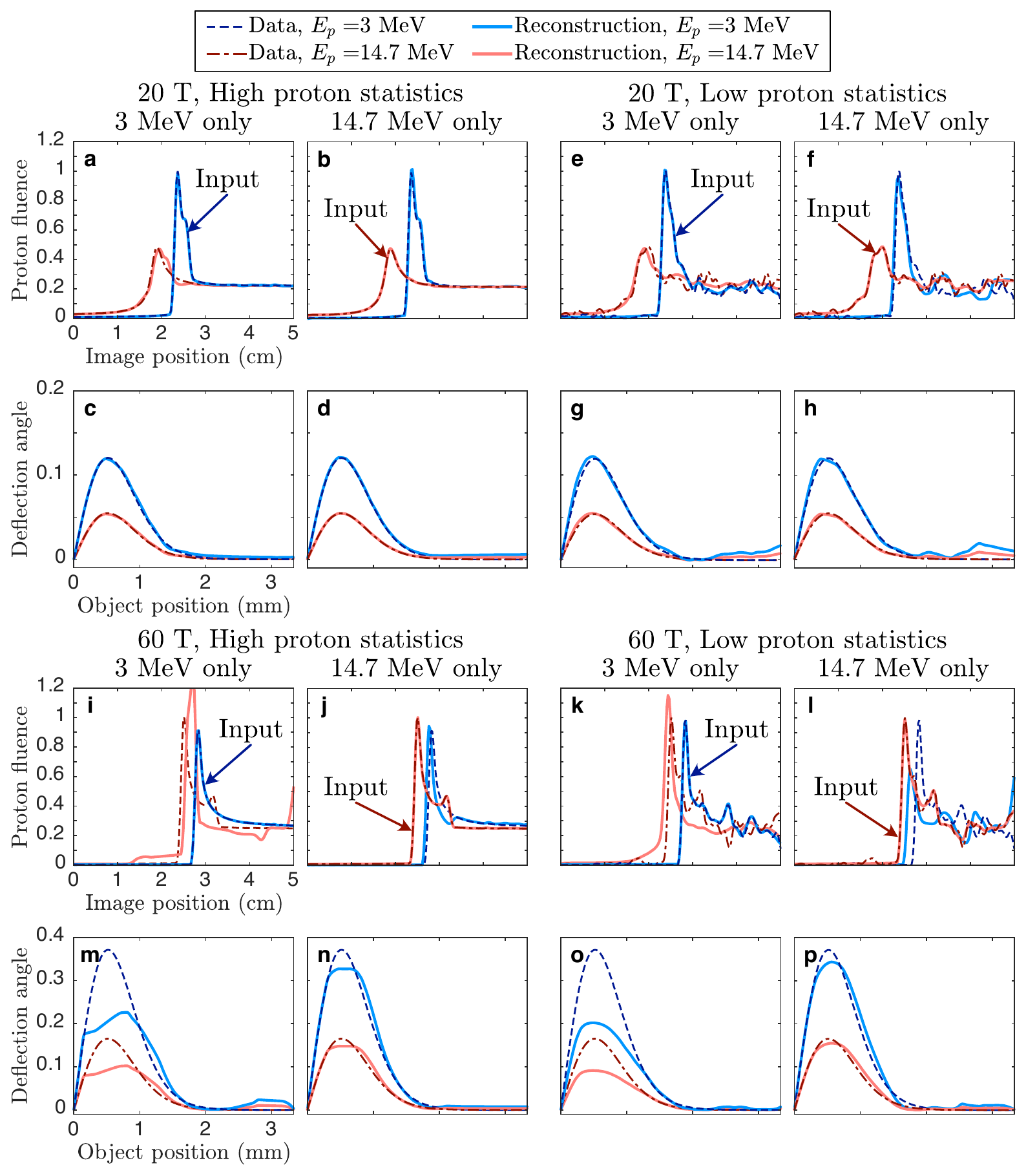}
	\caption{
	Reconstructing deflections for the Gaussian ellipsoidal field using only one input image, at 20 T (a-h), and 60 T (i-p), for the smooth (high proton statistics) and noisy (low proton statistics) images, showing both the reconstructed proton fluence and deflection angle.
	One image is sufficient for either initialization at 20 T.
	Only the 14.7 MeV proton image produces a suitable reconstruction at 60 T, when strong caustics are present.
	}
	\label{fig:EllipsoidRecon1Image}
\end{figure}

\begin{figure}
	\centering
	\includegraphics{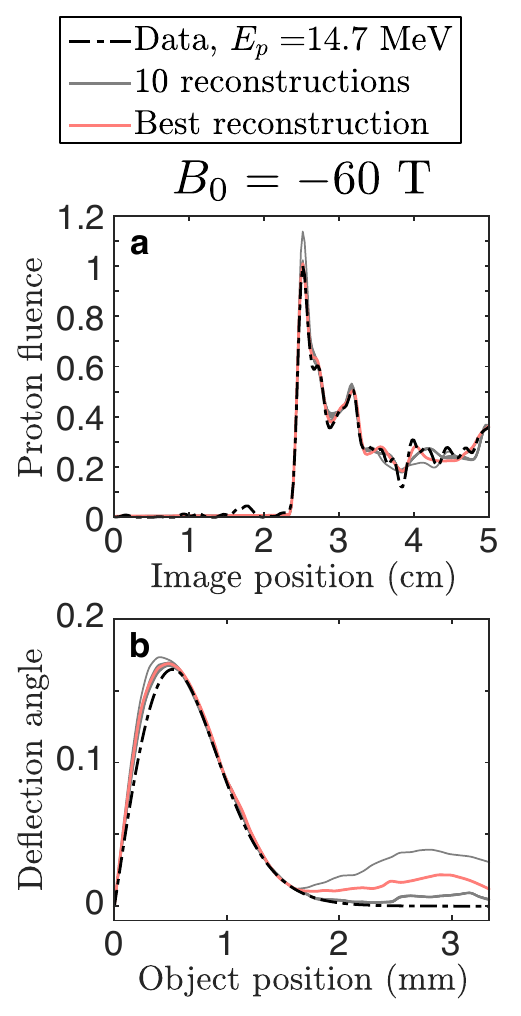}
	\caption{
	Reconstruction results from ten independent runs for the 60 T ellipsoidal field, showing only the 14.7 MeV portion for (a) the proton fluence profile, and (b) the deflection field.
	Most of the solutions (gray lines) fall within the same general profile, and deviate only slightly from the solution with the lowest error (red).
	}
	\label{fig:EllipsoidStatistics}
\end{figure}

To further demonstrate the robustness of this reconstruction technique, we apply it to images from the same Gaussian ellipsoidal field, but with fewer protons per image, reducing the proton statistics and introducing Poisson noise to the images.
The number of randomly initialized protons for the 2D images is reduced from $2\times 10^7$, as in Figure \ref{fig:EllipsoidRecon}, to only $10^5$ across the entire two-dimensional image, and line-outs of the proton intensity are averaged over a $\pi/40$ slice of the image.
To ensure that the Poisson noise between the two images is not correlated, the 1D intensity maps are taken from different sections of the image.
Figure \ref{fig:EllipsoidReconNoise}(a-c) shows the noisy intensity profiles for the same field strengths as Figure \ref{fig:EllipsoidRecon}, which display obvious variations across the image.
Despite the noise, it is apparent that the reconstructed deflections from the 20 T images in Figure \ref{fig:EllipsoidReconNoise}(d) is very close to the actual deflection field within the ellipsoid, although it deviates from the actual deflection as it approaches the edge of the image.
This trend is generally seen in Figure \ref{fig:EllipsoidReconNoise}(e) and \ref{fig:EllipsoidReconNoise}(f) for the reconstructions of higher fields as well, although with more pronounced deviations from the true deflection around the maximum.
In terms of how well the images are reconstructed, we can see that although the reconstruction now fits the intensity variations, it generally matches the peaks well and mostly remains near the average intensity toward the edge of the image.
This robustness in the reconstruction in the presence of noise likely stems from the reinforcing information from using two images with uncorrelated noise.

To test this assumption, we also look at how well our reconstruction method performs when given only a single image.
Applying this to the high-proton-statistics (smooth) images of Figure \ref{fig:EllipsoidRecon}, we reconstruct the deflections from either the 3 MeV or 14.7 Mev input only, the results are displayed in Figure \ref{fig:EllipsoidRecon1Image}(a-d) for the 20 T ellipsoidal field.
It is immediately apparent that our reconstruction method is able to accurately reconstruct the deflection field in both cases, with minimal deviation from the true deflections.
The intensity profiles from the 14.7 MeV-only reconstruction match both energies well, but the 14.7 MeV image corresponding to the 3 MeV-only reconstruction has some additional oscillations around the peak, showing that small changes in the deflection can produce significant intensity features.
So, for a smooth deflection field and an image with good proton statistics our reconstruction method also performs well from a single image.

When we try this single-image reconstruction on the noisy 20 T images, the results are somewhat surprising.
Figure \ref{fig:EllipsoidRecon1Image}(e-h) shows that even with lower proton statistics, the reconstruction is still able to find a deflection field very similar to the true deflection.
When comparing the intensities, we see that the they are also fairly good, but are fit more to the variations than with two images, as expected.

The 20 T ellipsoidal field does not produce particularly strong caustics, however, and the 14.7 MeV image is still in the linear regime, so we also test single-image reconstruction for the 60 T ellipsoidal field, and the results are shown in Figure \ref{fig:EllipsoidRecon1Image}(i-p).
Even for high proton statistics, reconstruction from the low-energy image alone produces a deflection field with significant deviation from the true deflection.
In this case, information has been lost, as some features are now outside of the field of view of the image.
In contrast, reconstruction from the 14.7 MeV image produces deflections fairly close to the actual deflection field, and the corresponding 3 MeV intensity profile resembles the real intensity.
When reducing the proton statistics to produce noisier input, we again see that the secondary energy intensity profiles do not match as well, but the reconstructed deflections remain close to those from the high-statistics reconstructions.
These tests demonstrate a major pitfall for reconstructing fields from a single image: this algorithm can produce a near-perfect fit to an image from deflections that deviate significantly from the actual deflection profile.
The second image is crucial for properly constraining the reconstruction, particularly when strong caustics are present.

We also test the reproducibility of the reconstructions in the presence of noise.
Figure \ref{fig:EllipsoidStatistics} shows the results from 10 independent reconstructions of the noisy, 60 T input using both images for the reconstruction.
Only four distinct solutions are observed, which are all very similar to the reconstruction with the lowest error, particularly around the regions of stronger deflection.
This test shows that when reconstructing deflections from two images, the fitness landscape defined by our set of heuristics and weights leads to a robust and reproducible solution from DE from these test cases.

\subsection{A current-carrying wire}\label{ss:Wire}

\begin{figure}
	\centering
	\includegraphics{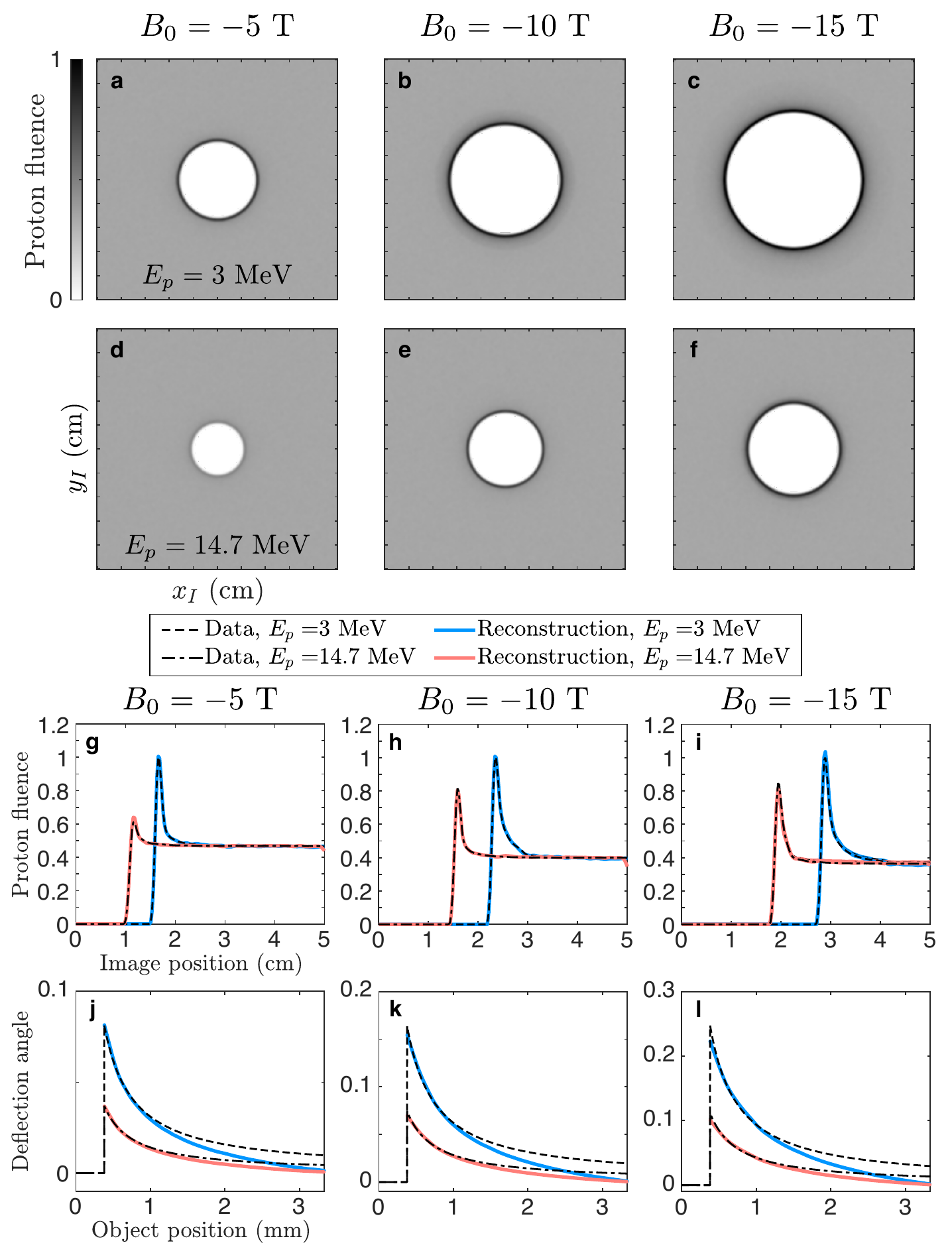}
	\caption{
	Synthetic proton images and reconstruction results from a simulated wire field as described by equation \ref{eq:WireField}, with $R=380$ $\mu$m at the magnetic field amplitudes $B_0 = -5$ T, $-10$ T, and $-15$ T.
	(a,b,c) Synthetic 3 MeV proton images.
	(d,e,f) Synthetic 14.7 MeV proton images.
	The sharp gradient of the field produces only a single caustic peak on each image.
	(g,h,i) The known proton intensity (dashed lines) and reconstructed proton intensity profiles (solid lines).
	(j,k,l) The known radial deflection field (dashed lines) and reconstructed deflections (solid lines).
	The proton fluence profiles are reconstructed well, but the deflection reconstruction deviates significantly after 1 mm.
	}
	\label{fig:WireRecon}
\end{figure}

Next we consider the proton images produced by the magnetic field around a straight current-carrying wire segment.
The field outside the wire is defined by
\begin{equation}\label{eq:WireField}
	B_{\phi}(r) = B_0\frac{R}{r},
\end{equation}
where $R$ is the inner radius of the solid wire, and $r$ is the distance from the center of the wire.
The magnetic field is again in $\phi$, so proton deflections are radially outward.
Figure \ref{fig:WireRecon}(a-f) shows the synthetic proton images for magnetic deflection from a 4 mm long segment of wire with $R=380$ $\mu$m, at $B_0=-5$ T, $-10$ T, and $-15$ T.
Proton images of the $1/r$ field profile configuration are inherently caustic at any significant field strength, and the resulting sharp intensity profiles make this a challenging field to reconstruct.

The results of our reconstruction for the wire field are shown in Figure \ref{fig:WireRecon}(g-i).
We initialize the reconstruction in the same way as for the Gaussian ellipsoid field, using the same number of nodes and the same stopping criteria.
The reconstructed proton fluence profiles again match the input very well, but the deflection field has significant deviations.
The reconstructed deflection fields match the actual deflection well for the first 1 mm of the object region, but decrease to zero much faster than the expected $1/r$.
The poor agreement between reconstructed and actual deflection here is likely because the images have single, sharp peaks followed by smooth, almost uniform fluence. 
These results suggest that in these cases there is only significant information to reconstruct the main driver of the caustic, while the rest of the image can be the result of much lower deflection, which is preferred by our chosen heuristics.

\subsection{A current-carrying wire with an exponential decrease}\label{ss:TWire}

\begin{figure}
	\centering
	\includegraphics{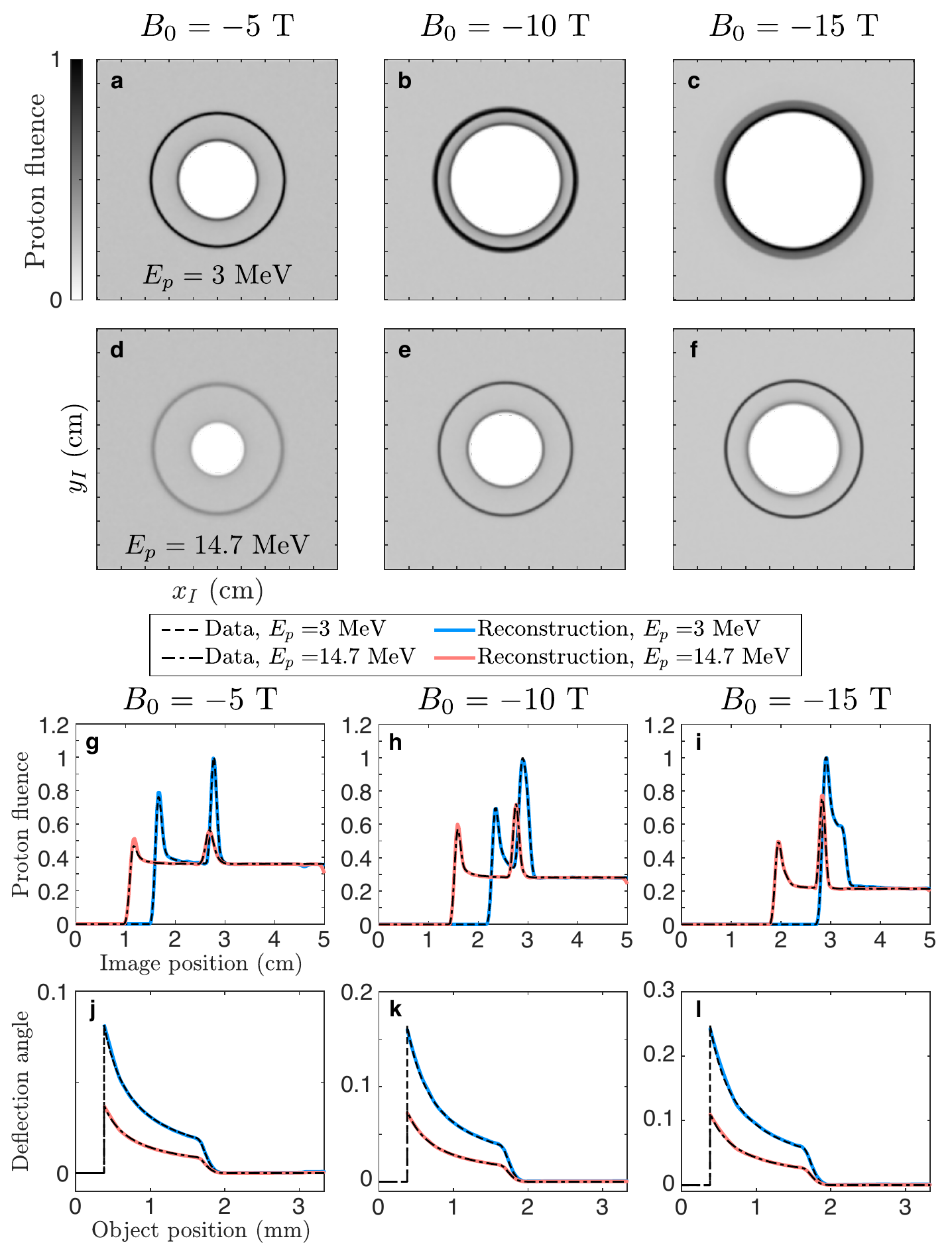}
	\caption{
	Synthetic test proton images and reconstruction results from a simulated current-carrying wire field with an additional Gaussian decrease as described by equation \ref{eq:TWireField}.
	(a,b,c) Synthetic 3 MeV proton images.
	(d,e,f) Synthetic 14.7 MeV proton images.
	Additional peaks are present because of the dropoff at a relatively fixed position regardless of field amplitude.
	(g,h,i) The known proton intensity (dashed lines) and reconstructed proton intensity profiles (solid lines).
	(j,k,l) The known radial deflection field (dashed lines) and reconstructed deflections (solid lines).
	With the additional information from the Gaussian reduction, our reconstruction method finds the underlying deflections almost exactly, in contrast to the wire-only field.
	}
	\label{fig:TWireRecon}
\end{figure}

Building on the previous section, we again set the base magnetic field from a current-carrying wire, but add a Gaussian decrease to zero with radius starting halfway across the object region, as
\begin{align}\label{eq:TWireField}
	B_{\phi}(r) = 
	\begin{cases}
    	B_0\frac{R}{r}\exp\left[-\left(\frac{r-1.62}{0.161}\right)\right]	& \text{if } r \ge 1.62\\
    	B_0\frac{R}{r}          & \text{\text{if } r < 1.62}
	\end{cases}
\end{align}
where the units of $r$ are again in mm.
Physically, this situation could correspond to the presence of a conducting medium (like a plasma) impinging on the wire field, but still allowing the transmission of protons with minimal scattering to produce an image.
The resulting proton images, shown in Figure \ref{fig:TWireRecon}, have additional features when compared to nominal wire field, which are strongly related to the curvature of the decrease.
The Gaussian decrease causes the deflection from the wire to suddenly stop, and creates another pileup of protons on the image --- a second caustic surface --- corresponding roughly to the position at which the decrease occurs.

The reconstruction results of Figure \ref{fig:TWireRecon} also show much better agreement than the nominal wire field case in both the proton fluence and deflection field with the input.
In fact, the deflection function is now consistently reconstructed almost exactly, not only is the Gaussian decrease recovered, but the preceding $1/r$ profile matches better than for the nominal wire field.
The increased fidelity is owed to the additional information from the second caustic on the image, which provides additional constraints on the deflection field.

\subsection{Application to experimental images}

\begin{figure}
	\centering
	\includegraphics{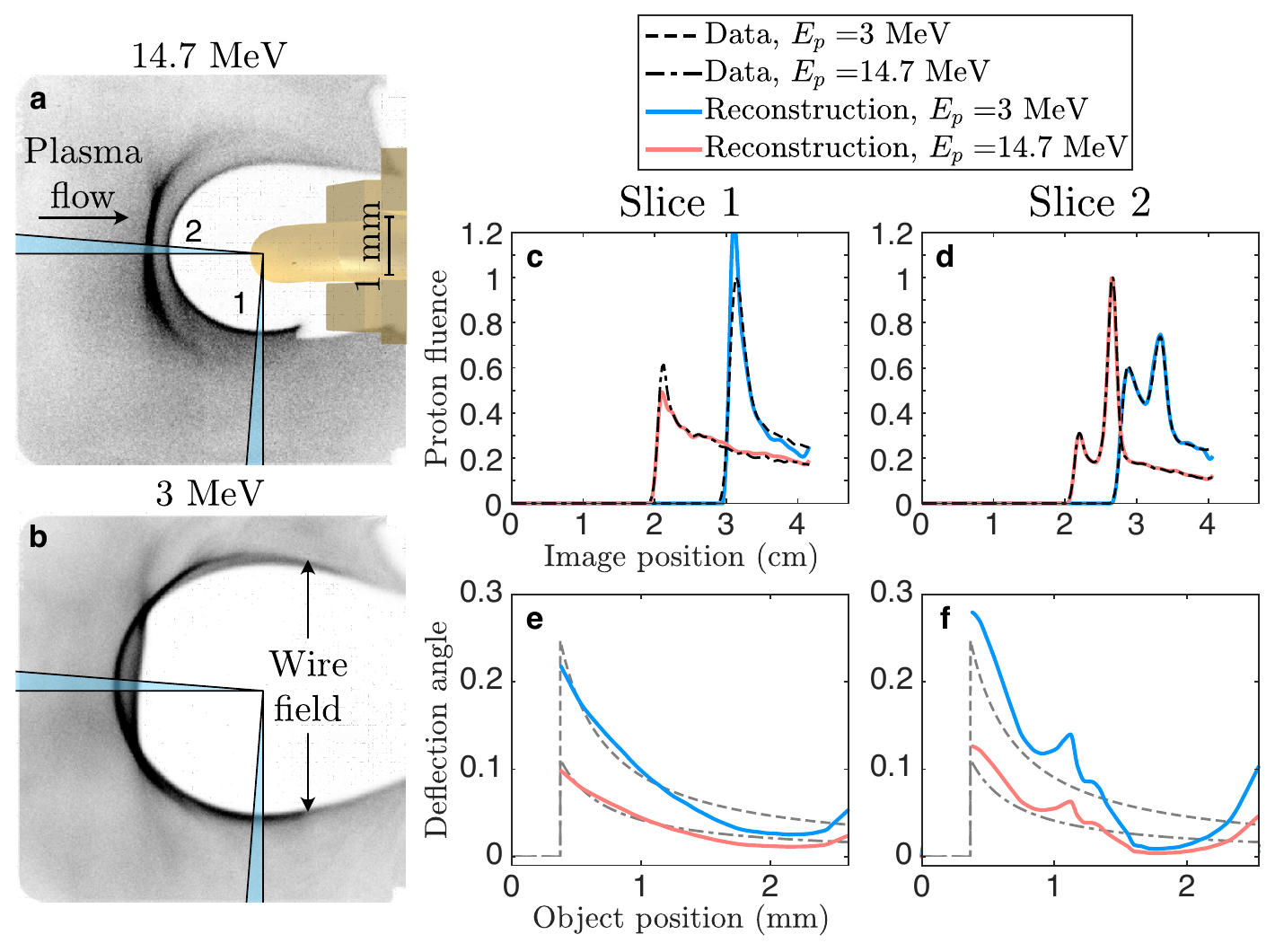}
	\caption{
	Applying the reconstruction method to experimental results.
	(a,b) The experimental images, with two slices outlined.
	(c,d) The input proton intensity (dashed lines) and reconstructed proton intensity profiles (solid lines) for the two slices.
	(e,f) The reconstructed deflections (solid lines), alongside the expected deflection from the magnetic field of wire segment at a maximum 13.5 T.
	}
	\label{fig:ExpRecon}
\end{figure}

For demonstration purposes, we also test our reconstruction method on a set of experimental proton images.
The images shown in Figure \ref{fig:ExpRecon}(a) and \ref{fig:ExpRecon}(b) are taken from an experiment looking at the interaction of an incoming plasma flow with the magnetic field generated by a current-carrying wire.
We run the reconstruction for the two highlighted slices of the overall image, each with an angular extent of $pi/40$.
The first slice is taken where there is little interference from the incoming plasma, so the field should be close to nominal.
The results in Figure \ref{fig:ExpRecon}(c) and \ref{fig:ExpRecon}(e) match the input proton fluence well, and the reconstructed deflection function resembles the deflections expected from a segment of current-carrying wire (from Section \ref{ss:Wire}) with a 13.5 T maximum field at the surface.
In reality, the wire has some finite curvature which makes direct comparison to a straight segment more difficult, but the reconstructed deflection appears to be about as good as the results from the synthetic tests in Figure \ref{fig:WireRecon}.
For the second slice, the incoming plasma is affecting the field structure, causing additional features on the proton image similar to that seen in Section \ref{ss:TWire}.
The reconstruction in Figure \ref{fig:ExpRecon}(d) and \ref{fig:ExpRecon}(f) fit the proton intensity almost exactly, and the deflection field shows a sharp increase around 1.2 mm from the assumed center of the wire.
The results we present use $w_1=1$, $w_2=2$, and $w_3=0.5$.
Unlike the test examples with smooth deflection fields, we find that the algorithm is very sensitive to $w_3$ for these images, and $w_3>1$ causes the reconstructed deflection to look very linear.
This extra sensitivity to $w_3$ suggests that additional loss of information from real experiments may require further tuning, or a search for alternative heuristics for improved performance.
The physical implications from this set of experiments will be explored in a future paper.

\section{Discussion}\label{sec:Discussion}

Using two simultaneous proton images at different proton energies, our differential evolution algorithm can accurately reconstruct synthetic, symmetric deflection field profiles regardless of if caustics are present on the images, with the exception of the wire-only field. 
We find that, at least in these cylindrically symmetric test cases, additional features in the proton image result in a better fit, likely because there is less degeneracy in the possible solution space.
This method can also be used with a single proton image input, reducing the accuracy of the reconstruction but still producing adequate deflection estimates in our tests when using the high-energy proton image. 
When the proton intensity is noisy, a second image is necessary to sufficiently constrain the solution space and avoid overfitting to noise.
However, if only one noisy image were available, smoothing the image prior to reconstruction may reduce the overfitting. 
Smoothing the image first will of course introduce another source of possible error in the reconstruction, but in the case of very low proton statistics this would likely be preferable to fitting small-scale signal variations.

In terms of computational cost, for these test cases our method generally converges by 300,000 iterations and usually takes less than 20 minutes running the reconstruction as a MATLAB script on a single 3.8 GHz processor.
There may be ways to further improve the convergence, particularly when increasing the resolution of the reconstruction, perhaps in the form of additional or alternative heuristics to optimize.
Other heuristics could likely be employed which achieve similar, or even better results, but it is not obvious what they would look like.
The heuristics chosen for this paper do have some physical meaning, and accomplish some desired limitations on the shape of the deflection function in a way general to the test cases, though they were chosen after much trial-and-error testing.
One possibility may be the inclusion of a probabilistic method which uses DE to minimize the log-likelihood of the deflection with respect to the input images.
Likelihood-based inference is attractive because it could also provide confidence intervals for the estimated deflection function.

The requirement of multiple images is not a significant barrier to many experimental applications, because obtaining multiple, nearly simultaneous images of a system is common in HED experiments.
As discussed earlier, experiments at the OMEGA laser facility using the D$^3$He capsule implosion generate protons at 3 MeV and 14.7 MeV.\cite{Li_RSI2006,Li_PoP2009}
Experiments using a capsule-implosion proton source typically evolve slowly compared to the duration of the source, producing two images of the same field structure at two very distinct proton energies.
Alternatively, the target-normal sheath acceleration (TNSA) method of proton source production at laser facilities generates a large band of high-energy protons, and the resulting energies are discriminated by being captured at successive film slices to produce many images in a single shot.\cite{Hatchett_PoP2000}
The reconstruction method we have presented can be trivially extended to incorporate an arbitrary number of input images, and the additional information available from a potentially large set of images could produce more accurate reconstruction of complex field geometries than with two images from the capsule implosion source.
Many experiments which use TNSA have so far focused on its ability to probe fields in fast-evolving systems\cite{Gao_PoP2016,Palmer_PoP2019}, rather than mostly static objects, and direct application of our reconstruction method may not be possible on images from these systems.
To properly apply our method to TNSA images, additional work would likely need to be done to determine how best to correlate images between fast-evolving systems and also to determine if there is an optimal or necessary proton energy difference between images to properly constrain the reconstruction.

Although we have restricted this paper to only consider pseudo-one-dimensional deflection fields, it is important to consider possible extension to two-dimensional deflection fields.
The underlying method of using multiple images should also be valid for reconstructing the full path-integrated field, but in terms of the differential evolution algorithm a number of complications need to be addressed.
First, the addition of a second dimension necessitates an increase in the number of nodes needed to sufficiently define the deflection field, with a corresponding increase in the computational cost.
Second, if allowing for nonuniform, asymmetric images the choice of node position will also be important for caustic reconstruction, although a sufficiently fine grid could avoid this issue at a cost.
Third --- and probably most importantly --- the underlying deflection function will require at minimum two values: the deflection amount and the deflection direction.
The two-component deflection solution space should be more efficient than considering a full three-component field.
It is not obvious as to whether both quantities will need to be evolved simultaneously, or in an alternating manner.
Additional heuristics would be needed to constrain the expanded solution space introduced by the additional direction component.
Even with new heuristics it is also not yet clear if two images would be enough to adequately constrain both the direction and the amount of deflection for reconstruction --- additional proton images may be required.

\section{Conclusion}\label{sec:Conclusion}

In conclusion, we have developed a new method for reconstructing underlying path-integrated magnetic deflections from a set of two proton images of the same object, each at a different proton energy.
We use a differential evolution algorithm to iteratively evolve an underlying deflection function to simultaneously reconstruct the proton fluence profiles on the two input images while minimizing a heuristic cost function.
The differential optimization scheme requires few assumptions in reconstruction, where most information comes from the presence of multiple images.
Our method is able to accurately reconstruct the deflections regardless of nonlinearities (caustics) in the proton image, as long as the images provide sufficient information.
This is among the first reconstruction methods truly able to deal with caustics, albeit in a limited geometry, and is equally applicable to non-caustic features.
Further improvements should enable expansion to arbitrary 2D geometries, and the differential evolution optimization could enable transition to and improvements in neural-network-based reconstruction.

\section*{Data availability}
The MATLAB methods developed for this work are publicly available at the GitHub repository \url{https://github.com/lanl/MEnDER-1D}. 
Additional data may be made available upon reasonable request.

\section*{Acknowledgements}

This work was partially funded by the U.S. Department of Energy, through the NNSA Center of Excellence under cooperative agreement number DE-NA0003869, the NNSA-DP and SC-OFES Joint Program in HEDLP, grant number DE-NA0002956, the NLUF Program and Rice University, grant numbers DE-NA0002722 and DE-NA0002719.
This work was also supported by the U.S. Department of Energy through the Los Alamos National Laboratory. Los Alamos National Laboratory is operated by Triad National Security, LLC, for the National Nuclear Security Administration of U.S. Department of Energy (Contract No. 89233218CNA000001).

\bibliography{References}

\end{document}